\documentstyle[preprint,prc,aps,psfig]{revtex}

\renewcommand{\vec}[1]{\mbox{\boldmath $#1$}}

\tightenlines

\begin{document}
\title{Role of dynamical particle-vibration coupling in 
reconciliation of the $d_{3/2}$ puzzle for spherical proton emitters} 
\author{K. Hagino}
\address{Yukawa Institute for Theoretical Physics, Kyoto
University, Kyoto 606-8502, Japan }

\maketitle

\begin{abstract}

It has been observed that decay rate for proton emission 
from $d_{3/2}$ single particle state is systematically 
quenched compared with the prediction of a one dimensional potential 
model although the same model successfully accounts for measured decay 
rates from $s_{1/2}$ and $h_{11/2}$ states. 
We reconcile this discrepancy by solving 
coupled-channels equations, 
taking into account 
couplings between the proton motion and vibrational excitations of a
daughter nucleus. 
We apply the formalism to proton emitting nuclei $^{160,161}$Re to 
show that there is a certain range of parameter set of the excitation 
energy and the dynamical deformation parameter for the quadrupole 
phonon excitation 
which reproduces simultaneously the experimental decay rates 
from the 2$d_{3/2}$, 3$s_{1/2}$ and 1$h_{11/2}$ states in these nuclei. 

\end{abstract}


Physics of nuclei close to the neutron and proton drip lines is one of the most
active and exciting research areas of the current nuclear physics. 
Nuclei beyond the proton drip line are unstable against proton emission,
but, since a proton has to penetrate the Coulomb barrier, their
lifetime is sufficiently long to study their spectroscopic
properties. 
Thanks to the recent experimental developments of production and
detection methods, a number of ground-state as well as isomeric 
proton emitters have recently been discovered, which has stimulated many
experimental and theoretical works 
\cite{procon99,WD97,PWC92,IDW97,DWB97,DWS98,BBR98,RBB99,BBR99,SDW99,ASN97,KBNV00,BKNV00,MFL98,FM00,MF00,FM01,DE00,ED00}. 

For proton emitters in the $A\sim$ 150 region, proton emissions 
from the 1$h_{11/2}$, 3$s_{1/2}$, and 2$d_{3/2}$ orbitals have been
observed. It has been pointed out that a spherical calculation based
upon a one dimensional optical potential with spectroscopic factor
estimated in the BCS 
approximation systematically underestimates the measured decay
half-lives 
for proton emissions from the 2$d_{3/2}$ state, while the same model 
works well for emissions from the 1$h_{11/2}$ and 3$s_{1/2}$ states 
in both odd-Z even-N nuclei and odd-Z odd-N nuclei 
\cite{DWB97,BBR99}. For the $^{151}$Lu nucleus, this discrepancy 
was attributed to the 
effects of oblate deformation of the core nucleus $^{150}$Yb, which
alter both the decay dynamics and the BCS occupation probability of
the valence proton \cite{BBR99,FM00,S01}. 
The coupled-channels calculations with $\beta_2 \sim -0.15$ 
have successfully explained the measured decay half-lives for
this nucleus \cite{FM00}. 
However, for proton emitters such as $^{156}$Ta, $^{160}$Re, and 
$^{161}$Re, the static deformation parameter of the core nuclei 
$^{155}$Hf, $^{159}$W, and $^{160}$W is estimated to be 
$\beta_2=-0.053$, 0.080, and 0.089, respectively, based on the
macroscopic-microscopic mass formula \cite{MNMS95}, and thus 
the deformation effects will be much smaller. 
Those nuclei are nearly spherical, and vibrational excitations should
be considered instead of the deformation effects and the associated
rotational excitations \cite{BKNV01}. 

For vibrational nuclei, the excitation energy of collective
excitations is in general significantly larger than that for
rotational nuclei. This makes the channel coupling effects 
weaker in spherical nuclei. We notice, however, that the penetration
probability is extremely sensitive to other degrees of freedom 
especially at deep barrier energies, regardless of the value of 
their excitation
energies \cite{HTDHL97,DLV91}. It is thus of interest and important
to explore the role of collective core excitations during proton
emission decays of spherical nuclei. 

The aim of this paper is to solve the coupled-channels equations for
spherical proton emitters in order to investigate 
whether the effects of vibrational 
excitations of the daughter nucleus consistently account for 
the measured decay half-life from 
the 1$h_{11/2}$, 3$s_{1/2}$, and 2$d_{3/2}$ states. 
We particularly study proton emissions from the 
1$h_{11/2}$ and 3$s_{1/2}$ states of $^{161}$Re \cite{IDW97} and 
from the 2$d_{3/2}$ state of $^{160}$Re \cite{PWC92}, assuming that 
the vibrational properties are identical between the core nuclei 
$^{160}$W and $^{159}$W. We shall exclude in our study the proton emitter 
$^{156}$Ta, since the experimental data is somewhat ambiguous due to 
the competition between the proton emission and the $\beta^+$/EC
decay. Since the properties of excited states of these proton-rich
nuclei are not known, we shall study the dependence of
the decay rate on the
excitation energy as well as on the dynamical deformation parameter of
the vibrational phonon excitation of the daughter nucleus. We will 
then extract 
a possible combination of these two from the experimental data. We will also
discuss the dependence on the multipolarity of the phonon state. 

We consider the following Hamiltonian for the spherical 
proton emitting systems: 
\begin{equation}
H=-\frac{\hbar^2}{2\mu}\nabla^2 +V_{coup}(\vec{r},\alpha)+H_{vib},
\end{equation}
where $\vec{r}$ is the coordinate for the relative motion
between the valence proton and the daughter nucleus, and $\mu$ the reduced
mass for this motion. $\alpha$ is the coordinate for the
vibrational phonon of the daughter nucleus. It is related to the
dynamical deformation parameter as 
$\alpha_{\lambda\mu}=\frac{\beta_{\mu}}{\sqrt{2\lambda+1}} 
(a_{\lambda\mu}^{\dagger}
+(-)^{\mu}a_{\lambda\mu})$, where $\lambda$ is the multipolarity of
the vibrational mode and $a_{\lambda\mu}^{\dagger} (a_{\lambda\mu})$ 
is the creation (annihilation) operator of the phonon. 
$H_{vib}=\hbar\omega \sum_{\mu}a_{\lambda\mu}^{\dagger}a_{\lambda\mu}$ 
is the Hamiltonian for the vibrational phonon. In this paper, for
simplicity, we do not consider the possibility of multi-phonon
excitations, but include only excitations to the single phonon state. 
The coupling Hamiltonian $V_{coup}(\vec{r},\alpha)$ consists of three
terms, {\it i.e.,} $V_{coup}(\vec{r},\alpha)=
V_{coup}^{(N)}(\vec{r},\alpha)+V_{coup}^{(ls)}(\vec{r},\alpha)+
V_{coup}^{(C)}(\vec{r},\alpha)$. The nuclear term reads 
\begin{equation}
V_{coup}^{(N)}(\vec{r},\alpha)= 
-\frac{V_0}
{1+\exp\left(\frac{r-R_0-R_0\,\alpha_{\lambda}\cdot
Y_{\lambda}(\hat{\vec{r}})}{a}\right)}\,,
\end{equation}
where the dot denotes a scalar product. We have assumed that the
nuclear potential is given by the Woods-Saxon form. 
Notice that we do not expand the nuclear potential but include the
couplings to the all orders with respect to the phonon operator 
$\alpha$ \cite{KRK99,HTDHL97b}. 
The matrix elements of this coupling Hamiltonian are evaluated using a
matrix algebra, as in Ref. \cite{KRK99}. As for the
Coulomb $V_{coup}^{(C)}$ as well as the spin-orbit $V_{coup}^{(ls)}$ 
terms, the effects of higher order couplings are expected to be 
small \cite{HTDHL97b}, and we retain only the linear term. 
The Coulomb term thus reads 
\begin{eqnarray}
V_{coup}^{(C)}(\vec{r},\alpha)&=& 
\frac{Z_D e^2}{r}+\frac{3Z_D e^2}{R_c}\,\frac{1}{2\lambda+1}
\left(\frac{R_c}{r}\right)^{\lambda}
\alpha_{\lambda}\cdot Y_{\lambda}(\hat{\vec{r}})~~~~~~~~~(for~r > R_c) \\
&=&\frac{Z_D e^2}{2R_c}\left(3-\frac{r^2}{R_c^2}\right)
+\frac{3Z_D e^2}{R_c}\,\frac{1}{2\lambda+1}
\left(\frac{r}{R_c}\right)^{\lambda}
\alpha_{\lambda}\cdot Y_{\lambda}(\hat{\vec{r}})~~~~~~~~~(for~r \leq R_c), 
\end{eqnarray}
where $Z_D$ is the atomic number of the daughter nucleus and $R_c$ is
the charge radius. 
For the spin-orbit interaction, we express it in the so called Thomas 
form \cite{ED00,S83}, 
\begin{equation}
V_{coup}^{(ls)}(\vec{r},\alpha)= 
V_{so}\frac{1}{r}\,\frac{df}{dr}\,\vec{l}\cdot\vec{\sigma}
+i\,V_{so}R_{so}\sum_{\mu}\alpha_{\lambda\mu}^*
\left\{\left(\nabla \frac{df}{dr}Y_{\lambda\mu}(\hat{\vec{r}})\right)
\cdot\left(\nabla \times \vec{\sigma}\right)\right\}\, , 
\label{ls}
\end{equation}
where $f(r)=1/[1+\exp((r-R_{so})/a_{so})]$. 
The last term in Eq. (\ref{ls}) 
can be decomposed to a sum of angular momentum tensors
using a formula \cite{E57}
\begin{eqnarray}
\left(\nabla g(r)Y_{\lambda\mu}(\vec{r})\right)\cdot \vec{C}
&=&-\sqrt{\frac{\lambda+1}{2\lambda+1}}\left(\frac{dg}{dr}-
\frac{\lambda}{r}\,g(r)\right)[Y_{\lambda+1}\vec{C}]^{(\lambda\mu)}
\nonumber \\
&&+\sqrt{\frac{\lambda}{2\lambda+1}}\left(\frac{dg}{dr}+
\frac{\lambda+1}{r}\,g(r)\right)[Y_{\lambda-1}\vec{C}]^{(\lambda\mu)}.
\end{eqnarray}

In order to solve the coupled-channels equations, we expand the total
wave function as 
\begin{equation}
\Psi_{jm}(\vec{r},\alpha)=\sum_{l_pj_p}\sum_{nI}
\frac{u^{(j)}_{l_pj_pnI}(r)}{r}\,|(l_pj_pnI)jm\rangle,
\end{equation}
where 
\begin{equation}
\langle \hat{\vec{r}},\alpha |(l_pj_pnI)jm\rangle =\sum_{m_p m_I}
\langle j_p m_p I m_I|jm\rangle {\cal Y}_{j_pl_pm_p}(\hat{\vec{r}})
\phi_{nIm_I}(\alpha), 
\end{equation}
$\phi$ being the vibrational wave function. 
We need to compute the coupling matrix elements of the operators 
$\alpha_{\lambda}\cdot T_{\lambda}$, where $T_{\lambda\mu}$ is either 
$Y_{\lambda\mu}$ or 
$[Y_{\lambda\pm 1}\,(-i \,\nabla \times \vec{\sigma})]^{(\lambda\mu)}$. 
These are expressed in terms of the Wigner's 6-j symbol as \cite{E57}
\begin{eqnarray}
\langle(l_p'j_p'n'I')jm|\alpha_{\lambda}\cdot T_{\lambda}
|(l_pj_pnI)jm\rangle &=&  
(-)^{j_p+I'+j}\left\{
\matrix{j & I' & j_p' \cr
\lambda & j_p & I \cr}\right\} \nonumber \\
&& \times \langle {\cal Y}_{j_p'l_p'}||T_{\lambda}||
{\cal Y}_{j_pl_p}\rangle \, 
\langle \phi_{n'I'}||\alpha_{\lambda}||\phi_{nI}\rangle. 
\end{eqnarray}
For transitions between the ground and the one phonon states which we
consider in this paper, the reduced matrix element 
$\langle \phi_{n'I'}||\alpha_{\lambda}||\phi_{nI}\rangle$ is given by 
$\beta_{\lambda}$. The reduced matrix elements for the operators 
$T_{\lambda}$ are found in Ref. \cite{RG92}. 

Our coupled-channels approach is based on the Gamow state wave function for
resonances. This indicates that the channel wave functions 
$u^{(j)}_{l_pj_pnI}(r)$ have the asymptotic form of 
$N^{(j)}_{l_pj_pnI}(G_{l_p}(k_{nI}r)+i\,F_{l_p}(k_{nI}r))$ at $r\to \infty$
for all the channels, 
where $k_{nI}=\sqrt{2\mu(E-n\hbar\omega)/\hbar^2}$ is the channel wave
number, and $F_{l_p}$ and $G_{l_p}$ are regular and
irregular Coulomb wave functions, respectively \cite{MFL98}. 
This method, however, requires to solve the
coupled-channels equations in the complex energy plane and out to
large distances, which is quite time consuming and also may be
difficult to obtain accurate solutions. 
A much simpler alternative approach has been proposed, which is based
on the Green's function formalism \cite{ASN97,DE00,ED00,K72,BK85,BK89}. 
This method was first developed for $\alpha$ decays by Kadmensky and 
his collaborators \cite{K72} and was recently applied to the 
coupled-channels problem for deformed proton emitters by 
Esbensen and Davids \cite{ED00}. 
In this method, the coupled-channels equations are solved in the real
energy plane and the solutions are matched to the irregular Coulomb
wave functions $G_{l_p}$ at a relatively small distance $r_{match}$, which is
outside the range of nuclear couplings. 
From the solution of the coupled-channels equations 
$\Psi^{cc}_{jm}(\vec{r},\alpha)$ thus obtained, 
the outgoing wave function
for the resonance Gamow state is generated using the Coulomb
propagator as \cite{DE00,ED00}
\begin{equation}
\Psi_{jm}(\vec{r},\alpha) = -\int d\vec{r}'d\alpha'
\left\langle\vec{r}\alpha \left|
\frac{1}{H_{coul}+H_{vib}-E-i\eta}\right|\vec{r}'\alpha'\right\rangle
\left(V_{coup}(\vec{r}',\alpha')-\frac{Z_De^2}{r'}\right)
\Psi^{cc}_{jm}(\vec{r}',\alpha'), 
\end{equation}
where $\eta$ is an infinitesimal number and 
$H_{coul}=-\hbar^2\nabla^2/2\mu+Z_De^2/r$ is the Hamiltonian for the
point Coulomb field. As is well known, the single particle Green's
function is expressed in terms of the regular and the outgoing wave
functions \cite{DE00}. The asymptotic normalization factors 
$N^{(j)}_{l_pj_pnI}$ then read \cite{ED00}
\begin{equation}
N^{(j)}_{l_pj_pnI}=-\frac{2\mu}{\hbar^2k_{nI}}\int^{\infty}_0 dr\,r 
F_{l_p}(k_{nI}r)\left\langle (l_pj_pnI)jm \left|
V_{coup}(\vec{r},\alpha)-\frac{Z_De^2}{r}\right|
\Psi^{cc}_{jm}\right\rangle.
\end{equation}
In this way, the effects of the long range Coulomb couplings outside
the matching radius $r_{match}$ are treated perturbatively. 
Computing the asymptotic outgoing flux, the total decay width 
is found to be 
\begin{equation}
\Gamma_j=\sum_{l_pj_pnI}\frac{\hbar^2k_{nI}}{\mu}|N^{(j)}_{l_pj_pnI}|^2\,
\frac{1}{\langle\Psi^{cc}_{jm}|\Psi^{cc}_{jm}\rangle}\, , 
\end{equation}
where the normalization factor 
$\langle\Psi^{cc}_{jm}|\Psi^{cc}_{jm}\rangle$ is calculated inside the
outer turning point. 
The decay half-life is then defined as 
\begin{equation}
T_{1/2}=\frac{\hbar}{S_j \Gamma_j}\ln 2,
\label{halflife}
\end{equation}
where $S_j$ is the spectroscopic factor for the resonance state. 
If one assumes that the ground state of an odd-Z nucleus is a
one-quasiparticle state, the spectroscopic factor $S_j$ is 
identical to the unoccupation probability for this state and is given by 
$S_j=u_j^2$ in the BCS approximation \cite{ASN97}. 

Let us now numerically solve the coupled-channels equations for 
the resonant 1$h_{11/2}$ and 3$s_{1/2}$ states in $^{161}$Re as well
as the 2$d_{3/2}$ state in $^{160}$Re. 
We use the real part of the Becchetti-Greenless optical model 
potential for the proton-daughter nucleus potential \cite{BG69}. 
The potential depth was adjusted so as to reproduce the
experimental proton decay $Q$ value for each value of the dynamical
deformation parameter $\beta$ and the excitation energy $\hbar\omega$
of the vibrational phonon excitations in the daughter nucleus. 
Following Ref. \cite{ASN97}, we assume that the depth of the spin-orbit 
potential is related to that of the central potential by  
$V_{so}=-0.2V_0$. The charge radius $R_c$ is assumed to be the same as
$R_0$ in the nuclear potential. 
The spectroscopic factor $S_j$ in Eq. (\ref{halflife}) is taken from Ref. 
\cite{ASN97}. This was evaluated in the BCS approximation to 
a monopole pairing Hamiltonian for 
single-particle levels obtained with a spherical Woods-Saxon
potential. 

We first discuss the effects of the quadrupole vibrational
excitations.  
Figure 1 shows the dependence of 
the decay half-life for proton emissions from the three resonance
states on the dynamical deformation parameter of the quadrupole mode. 
The excitation energy of the quadrupole phonon is set to be 0.8
MeV. The experimental values for the decay half-life 
are taken from Refs. \cite{PWC92,IDW97}, 
and denoted by the dashed lines. The arrows are the half-lives in the
absence of the vibrational mode. For the decays from the 3$s_{1/2}$
and 1$h_{11/2}$ states, if one takes into account uncertainty in the
experimental $Q$ value of the proton emission, these calculations 
for the decay half-life in the no coupling limit are within the 
experimental error bars \cite{ASN97}. 
In contrast, the calculation for the 2$d_{3/2}$ state is still off
from the experimental data by about 30\% even when uncertainty of the
decay $Q$ value is taken into consideration \cite{ASN97}. 
The results of the coupled-channels calculations are shown by
the solid lines in the figure. One notices that the channel coupling
effects significantly enhance the decay half-life for the 
2$d_{3/2}$ state, while the effects are more marginal for the 
3$s_{1/2}$ and 1$h_{11/2}$ states (notice the difference of the scale
of the vertical axis 
in the figure). If one assumes that the vibrational properties are
identical between the core 
nuclei $^{159}$W and $^{160}$W, the dynamical deformation parameter of
the range 0.18 $\leq \beta_2 \leq$ 0.23 
simultaneously reproduces the measured decay half-life for the 
2$d_{3/2}$, 3$s_{1/2}$ and 1$h_{11/2}$ states. 

Figure 2 shows the results for different value of the excitation
energy of the vibrational phonon. For each panel, with the decreasing
order, the four curves denote the decay half-life for
$\hbar\omega$=0.6, 0.7,0.8, and 0.9 MeV, respectively. 
The half-life for the 3$s_{1/2}$ state is relatively sensitive to the
phonon energy. For the phonon energy smaller than about 0.6 MeV, the curve 
raises too quickly as a function of the dynamical deformation
parameter, and the consistent description among the decay rates for the
three states is not possible. The range of dynamical
deformation parameter which simultaneously reproduces the measured
decay half-life for the three resonance states is plotted in
fig. 3 as a function of the excitation energy $\hbar\omega_2$. 
We see that the simultaneous description is possible only when 
$\hbar\omega_2$ is larger than 0.59 MeV. The value of $\beta_2$ larger
than 0.25 would not be acceptable for spherical nuclei, but the
minimum value of $\beta_2$ is always below this limit for $\hbar\omega
\ge$ 0.6 MeV. 

Lastly, we would like to discuss the dependence on the multipolarity
of the phonon mode. Figure 4 shows the effects of octupole phonon
excitations on the decay half-life from the 2$d_{3/2}$ state of 
$^{160}$Re. As an illustrative example, we take $\hbar\omega_3$=0.2
MeV, but results are qualitatively the same for different values of 
$\hbar\omega_3$. For the octupole vibration, the enhancement of the
half-life is too small to account for the observed discrepancy between
the experimental decay half-life and the prediction of the potential model
with no coupling. As was noted before \cite{WD97}, the proton decay is
very sensitive to the angular momentum of the proton state. The same
seems to be true for the multipolarity of the collective excitation of
the daughter nucleus. 

In summary, we have solved the coupled-channels equations to take into
account the effects of the vibrational excitations of the daughter
nucleus during the proton emission. Applying the formalism to the 
the resonant 1$h_{11/2}$ and 3$s_{1/2}$ states in $^{161}$Re and 
the 2$d_{3/2}$ state in $^{160}$Re, we have found that the
experimental data for the decay half-lives for these three states 
can be reproduced if the quadrupole phonon excitation with
$\hbar\omega_2 \ge$ 0.6 MeV and $\beta_2 \sim 0.18$ is considered. 
This removes the discrepancy observed before between the experimental
data and the prediction of the optical potential model calculation for
the decaying 2$d_{3/2}$ state in this mass region without destroying
the agreement for the 1$h_{11/2}$ and 3$s_{1/2}$ states. 
A similar calculation with the octupole phonon was not satisfactory. 
In this paper, we estimated the spectroscopic factor in the BCS
approximation for the monopole pairing Hamiltonian. It would be an
interesting future work to consistently compute both the decay rate 
with the coupled-channels framework and the  
spectroscopic factor based on the phonon induced pairing mechanism
proposed in Ref. \cite{BBG99}.

\newpage

\begin{figure}
  \begin{center}
    \leavevmode
    \parbox{0.9\textwidth}
           {\psfig{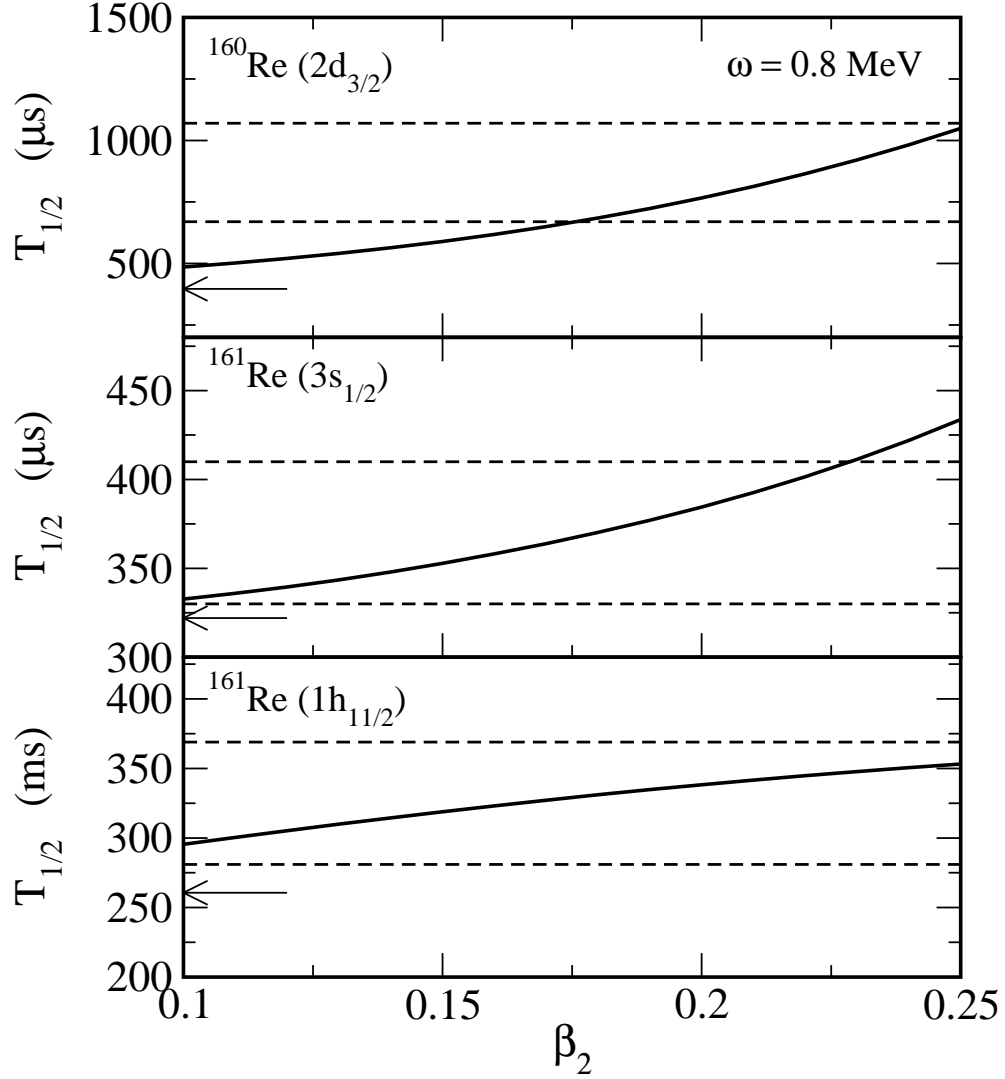}}
  \end{center}
\protect\caption{
The decay half-lives for proton emission from the 2$d_{3/2}$ 
state of $^{160}$Re and the 
3$s_{1/2}$ and 1$h_{11/2}$ states of $^{161}$Re as a function of the 
dynamical deformation parameter $\beta_2$ of the quadrupole
vibrational excitation of the daughter nuclei. The excitation energy 
of the quadrupole phonon is set to be 0.8 MeV.  
The experimental data are taken from Refs. [3,4] and
denoted by the dashed lines. 
The arrows indicate the results in the no coupling limit.} 
\end{figure}

\newpage

\begin{figure}
  \begin{center}
    \leavevmode
    \parbox{0.9\textwidth}
           {\psfig{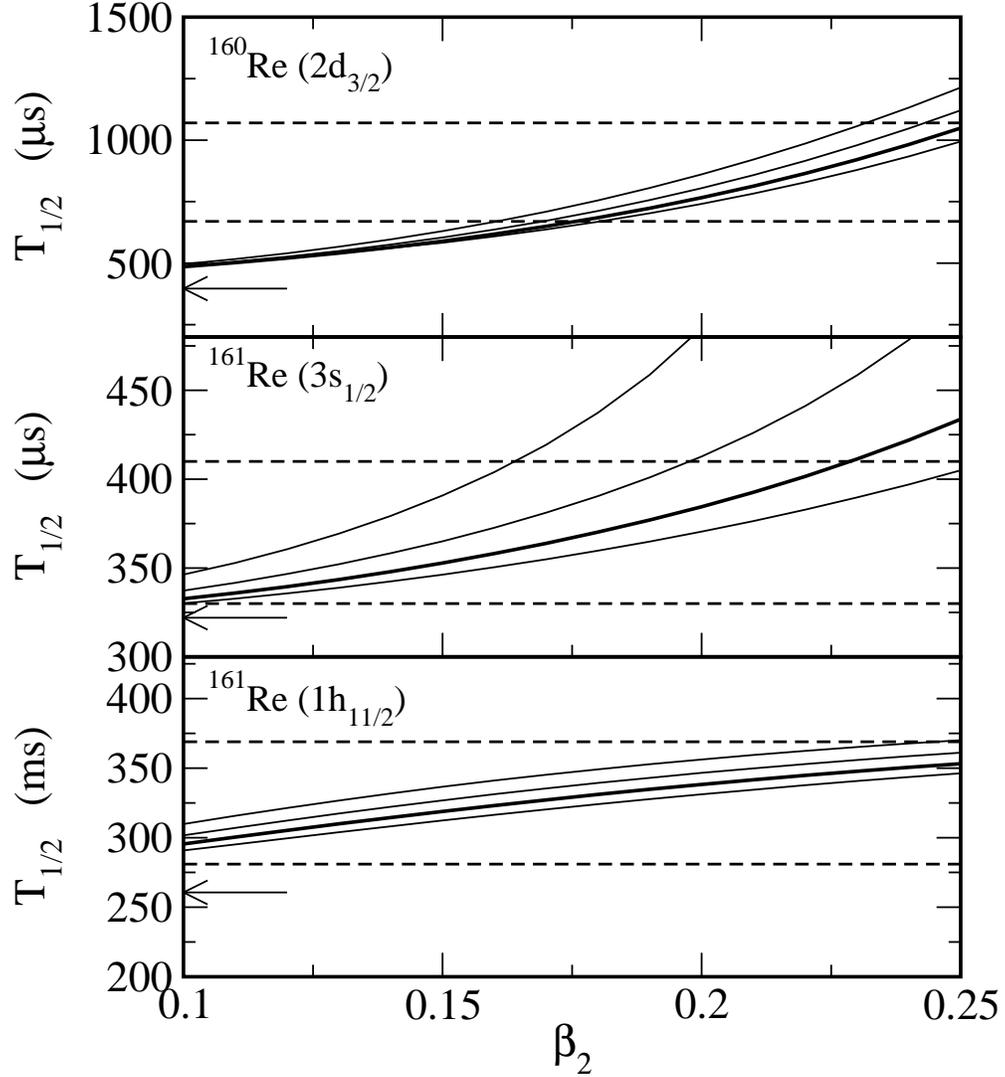}}
  \end{center}
\protect\caption{
The dependence of the half-lives on the value of the excitation energy
of the quadrupole phonon. In the decreasing order, the curves are for 
$\hbar\omega_2$ = 0.6, 0.7, 0.8, and 0.9 MeV, respectively. The
meaning of each line is the same as in Fig. 1. } 
\end{figure}

\newpage
\begin{figure}
  \begin{center}
    \leavevmode
    \parbox{0.9\textwidth}
           {\psfig{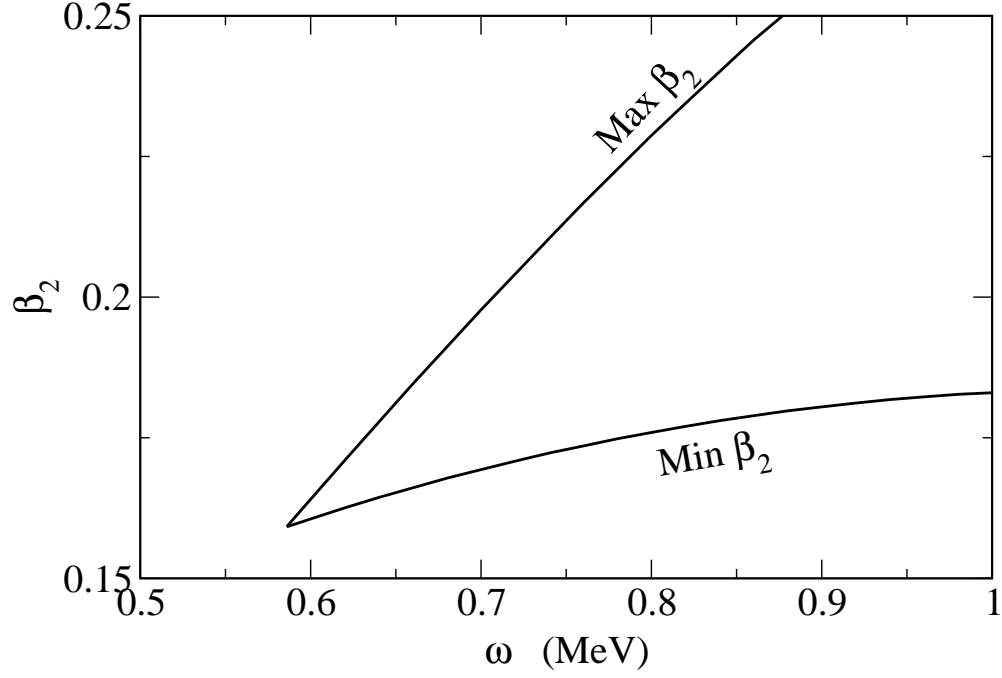}}
  \end{center}
\protect\caption{
The range of the dynamical deformation parameter $\beta_2$ and the
excitation energy $\hbar\omega$ which simultaneously reproduces the
experimental decay half-life for 
the 2$d_{3/2}$ state of $^{160}$Re and the 
3$s_{1/2}$ and 1$h_{11/2}$ states of $^{161}$Re. }
\end{figure}

\newpage

\begin{figure}
  \begin{center}
    \leavevmode
    \parbox{0.9\textwidth}
           {\psfig{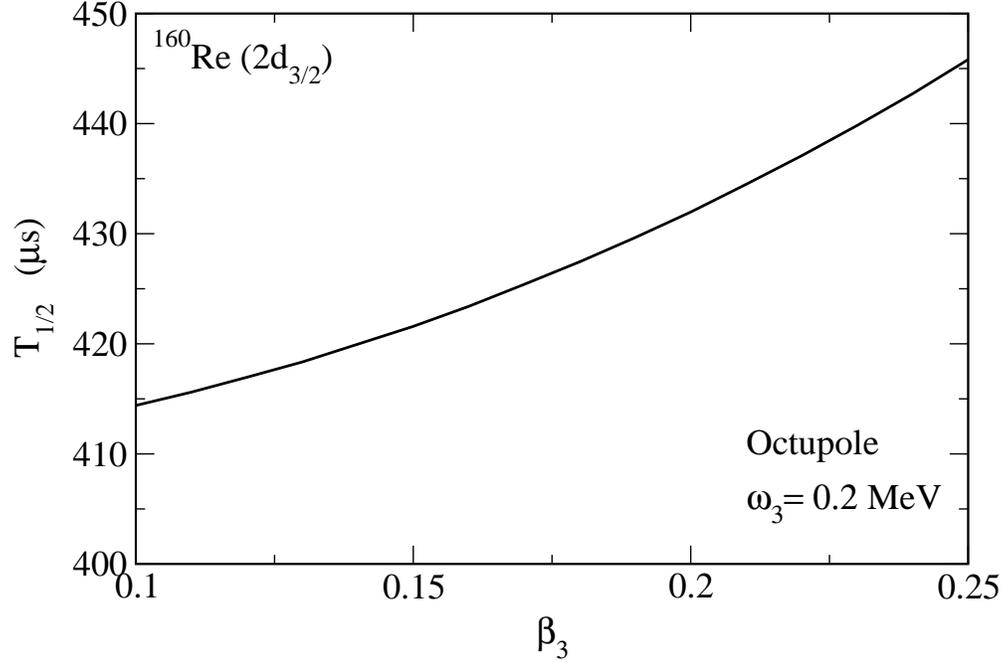}}
  \end{center}
\protect\caption{
The decay half-life for proton emission from the 
2$d_{3/2}$ state of $^{160}$Re as a function of the dynamical
deformation parameter of the octupole phonon excitation of the core
nucleus. The excitation energy is set to be 0.2 MeV.} 
\end{figure}

\end{document}